# HUBO and QUBO models for Prime factorization

Kyungtaek Jun[1,*]

1 Institute of Mathematical Sciences, Ewha Womans University, South Korea

∗Corresponding author: ktfriends@gmail.com

**Abstract**

The security of the RSA cryptosystem is based on the difficulty of factoring a large number $N$ into prime numbers $p$ and $q$ satisfying $N = p \times q$. This paper presents a prime factorization method using D-Wave quantum computer that can threaten the RSA cryptosystem. The starting point for this method is very simple, representing two prime numbers as qubits. Then, set the difference between the product of two prime numbers expressed in qubits and N as a cost function, and find the solution when the cost function becomes the minimum. D-Wave's quantum annealer can find the minimum value of any quadratic problem. However, the cost function is to be a higher-order unconstrained optimization (HUBO) model because it contains the second or higher order terms. We used a hybrid solver and *dimod* package provided by D-Wave Ocean software development kit (SDK) to solve the HUBO problem. We also successfully factorized 102,454,763 with 26 logical qubits. In addition, we factorized 1,000,070,001,221 using the range dependent Hamiltonian algorithm.

1. **Introduction**

  RSA cryptosystem is a popular public-key cryptographic algorithm for secure data transmission, first introduced in 1978 [1]. Public and private keys form a pair in the RSA cryptosystem. Content encrypted with the public key can only be decrypted with the private key, and content encrypted with the private key can only be decrypted with the public key. The public key is a large bi-prime that can only be decrypted through prime factors held in the private key. The RSA cryptosystem is based on prime factorization [2], which finds the prime factors $p$ and $q$ such that $N = p \times q$ for a large bi-prime $N$. The algorithm is an interesting mathematical problem because the algorithm relies on principles of number theory. While the RSA cryptosystem is surprisingly simple, a large bi-prime is difficult to factorize [3]. In other words, the computational complexity of decryption is much higher than that of encryption, so security is excellent. However, with the development of quantum computers, quantum algorithms that reduce the computational complexity difference between encryption and decryption are proposed, threatening the safety of the RSA cryptosystem.

  Peter Shor proposed the Shor algorithm [4], a powerful quantum algorithm that can factorize an integer $N$ in polynomial time, so can attack the RSA cryptosystem in polynomial time in 1994 [5]. There have been many simulations on quantum computers for practical use—moreover, many attempts to run the Shor algorithm on quantum computer hardware. Geller et al. applied the Shor algorithm to factorize 51 and 85 using Fermat numbers and eight qubits [6]. However, there are still limitations to applying Shor's algorithm to a large number. On the other hand, quantum annealing, which underlies a D-Wave's quantum computer, allows the factorization of even larger numbers. D-Wave's quantum annealing can find the minimum value of the quadratic unconstrained binary optimization (QUBO) or Ising model [7]. The objective formulated as the problem Hamiltonian of a D-Wave quantum computer is achieved by defining the linear and quadratic coefficients of a binary quadratic model (BQM) that maps those values to the qubits and couplers of the QPU. Dridi et al. presented a new autonomous algorithm that factorizes all bi-primes up to 200,099 by reducing Hamiltonian's degree using Grobner bases [8]. Jiang et al. developed a frame that transforms the prime factorization problem for arbitrary integers into the Ising model using ancillary variables and factorized the integer 376,289 with 94 logical qubits via a D-Wave 2000Q System [9]. Due to the limitation of the number of usable qubits of the D-Wave 2000Q System, many researchers have studied factorizing of a larger integer using the D-Wave's hybrid quantum/classical simulator *qbsolv*. Peng et al. improved Jiang et al.'s work and factored 1,005,973 with 89 qubits by using *qbsolv* [10]. Wang et al. successfully factorized all integers within 10,000 and demonstrated 1,028,171 with 88 qubits via *qbsolv* [11]. Also, Wang et al. recently factorized 1,630,729 (an 11-bit prime factor multiplied by an 11-bit prime factor) [12]. Unfortunately, a generalized method for factoring all integers into primes has not been developed due to the limitations of current quantum computers.

Jiang et al. (2018) proposed a direct method for prime factorization. In this paper, we propose a direct method to formulate a higher-order unconstrained optimization (HUBO) model for prime factorization. To obtain the HUBO model, it is sufficient to express $p$ and $q$, which are prime factors of bi-prime N, as sums of qubits, respectively. The minimum value of the HUBO model can be obtained via a Leap's quantum-classic hybrid solver provided by the D-Wave Ocean software development kit (SDK) [13]. The hybrid solver is an updated version of *qbsolv*. The hybrid solver can solve arbitrary problems formulated as quadratic models. Therefore, a process of converting the HUBO model to the QUBO model is required to solve a problem using a hybrid solver. The composed sampler in the D-Wave Ocean's *dimod* package creates a BQM from a higher order problem. Hence, if these are used, prime factorization of large integers is possible using the D-Wave Ocean SDK. We demonstrated the factorization of 102,454,763 with 26 logical qubits using the D-Wave hybrid solver as a successful example of our method. In addition, we applied the range dependent Hamiltonian algorithm that divides the domain to the new HUBO model. This algorithm is a method of finding a solution with a small number of qubits by dividing the domain into certain subintervals. The number of 1,000,070,001,221 was prime factorized using the HUBO model to which this algorithm was applied.

## 2. Methods

### 2.1. The least-squares problem for prime factorizations

The Ising model is a mathematical model for ferromagnetism in statistical mechanics. The energy Hamiltonian (the cost function) is formulated as follows:

$$H(\vec{\sigma}) = -\sum_{i=1}^{N} h_i \sigma_i - \sum_{i<j}^{N} J_{i,j} \sigma_i \sigma_j \tag{1}$$

where $\vec{\sigma} = (\sigma_1, \cdots, \sigma_N)^T$ and $\sigma_i \in \{+1, -1\}$.

QUBO is a combinatorial optimization problem in computer science. In this problem, a cost function $f$ is defined on an $n$-dimensional binary vector space $\mathbb{B}^n$ onto $\mathbb{R}$.

$$f(\vec{q}) = \vec{q}^T Q \vec{q} \tag{2}$$

where $Q$ is an upper diagonal matrix, $\vec{q} = (q_1, \cdots, q_N)^T$, and $q_i$ is a binary element of $\vec{q}$. In this paper, the matrix Q is referred to as the QUBO matrix. The problem is to find $\vec{q^*}$ which minimizes the cost function $f$ among vectors $\vec{q}$. Since we have $q_i^2 = q_i$, the cost function is reformulated as follows:

$$f(\vec{x}) = \sum_{i=1}^{N} Q_{i,i} q_i + \sum_{i<j}^{N} Q_{i,j} q_i q_j \tag{3}$$

In $Q$, the diagonal terms $Q_{i,i}$ and the off-diagonal terms $Q_{i,j}$ represent the linear terms and the quadratic terms, respectively. The unknowns of the Ising model $\sigma$ and the unknowns of the QUBO model $q$ have the linear relation

$$\sigma \to 2q - 1 \text{ or } q \to \frac{1}{2}(\sigma + 1) \tag{4}$$

Assume that the integer $N$ is the product of two prime numbers $p$ and $q$. To calculate $p$ and $q$, consider the least square problem as below:

$$\arg\min_{p,q} \| pq - N \| \tag{5}$$

Equation (1) satisfies the minimum value 0 when $pq = N$. To compute Eq. (5) conveniently, we apply 2-norm square to it.

$$\| pq - N \|_2^2 = p^2q^2 - 2pqN + N^2 \tag{6}$$

### 2.2. HUBO model

While solving the binary least-squares problem, $p$ and $q$ are represented by combinations of qubits $q_l \in \{0,1\}$. The radix 2 representation of the positive integer values, $p$ and $q$, given by

$$p \approx \sum_{l=0}^{n-1} 2^l q_l \text{ and } q \approx \sum_{l=0}^{n-1} 2^l q_{n+l} \tag{7}$$

where the integer, $l+1$, denotes the number of binary digits of $p$ and $q$ [14]. We use qubits from $l = 0$ to use the same equation in the range dependent Hamiltonian algorithm in Chapter 2.4.

To derive a HUBO model, we insert Eq. (7) into Eq. (6). This yields the summation terms of the first term in Eq. (6), as indicated below:

$$p^2q^2 = \left(\sum_{l=0}^{n-1} 2^l q_l\right)^2 \left(\sum_{l=0}^{n-1} 2^l q_{n+l}\right)^2 \tag{8}$$

$$= \left(\sum_{l=0}^{n-1} 2^{2l} q_l + \sum_{l_1 < l_2} 2^{l_1 + l_2 + 1} q_{l_1} q_{l_2}\right) \left(\sum_{l=0}^{n-1} 2^{2l} q_{n+l} + \sum_{l_1 < l_2} 2^{l_1 + l_2 + 1} q_{n+l_1} q_{n+l_2}\right) \tag{9}$$

$$= \sum_{l_1=0}^{n-1} \sum_{l_2=0}^{n-1} 2^{2(l_1+l_2)} q_{l_1} q_{n+l_2} + \sum_{l_1=0}^{n-1} \sum_{l_2 < l_3} 2^{2l_1 + l_2 + l_3 + 1} \left(q_{l_1} q_{n+l_2} q_{n+l_3} + q_{l_2} q_{l_3} q_{n+l_1}\right) \tag{10}$$

$$+ \sum_{l_1 < l_2} \sum_{l_3 < l_4} 2^{l_1 + l_2 + l_3 + l_4 + 2} q_{l_1} q_{l_2} q_{n+l_3} q_{n+l_4}$$

In Eq. (8), since $(q_l)^2 = q_l$, equation (9) can be obtained. The second term of Eq. (6) is calculated as follows:

$$-2pqN = -2N \left(\sum_{l=0}^{n-1} 2^l q_l\right) \left(\sum_{l=0}^{n-1} 2^l q_{n+l}\right) \tag{11}$$

$$= \sum_{l_1=0}^{n-1} \sum_{l_2=0}^{n-1} \left(-2^{l_1 + l_2 + 1} N q_{l_1} q_{n+l_2}\right) \tag{12}$$

The HUBO model consists of the sum of Eqs. (10) and (12). And the global minimum energy we need to get is $-N^2$.

### 2.3. QUBO model

The HUBO model for prime factorizations consists of quadratic, cubic, and quartic terms. To reformulate a non-quadratic (higher degree) polynomial into QUBO form, terms of the form, $cxyz$, where $c$ is a real number, are substituted with one of the following quadratic terms [9]:

$$cxyz = \begin{cases} cw(x + y + z - 2), & c < 0 \\ c\{w(x + y + z - 1) + (xy + yz + zx) - (x + y + z) + 1\}, & c > 0 \end{cases} \tag{13}$$

For all $x, y, z \in \{0,1\}$, $cxyz$ can be transformed into a combination of linear and quadratic terms by adding a new qubit $w$ to every cubic term. Similarly, Eq. (13) can be applied twice to convert quartic terms into QUBO formulations. The quartic terms with positive coefficients of this HUBO model are calculated as follows:

$$a_1 a_2 b_1 b_2 = a_1(x_1 a_2 + x_1 b_1 + x_1 b_2 + a_2 b_1 + a_2 b_2 + b_1 b_2 - a_2 - b_1 - b_2 - x_1 + 1) \tag{14}$$

$$\begin{aligned}
&= x_2 x_1 + x_2 a_1 + x_2 a_2 + x_1 a_1 + x_1 a_2 + a_1 a_2 - x_1 - a_1 - a_2 - x_2 + 1 \\
&+ x_3 x_1 + x_3 a_1 + x_3 b_1 + x_1 a_1 + x_1 b_1 + a_1 b_1 - x_1 - a_1 - b_1 - x_3 + 1 \\
&+ x_4 x_1 + x_4 a_1 + x_4 b_2 + x_1 a_1 + x_1 b_2 + a_1 b_2 - x_1 - a_1 - b_2 - x_4 + 1 \\
&+ x_5 a_1 + x_5 a_2 + x_5 b_1 + a_1 a_2 + a_1 b_1 + a_2 b_1 - a_1 - a_2 - b_1 - x_5 + 1 \\
&+ x_6 a_1 + x_6 a_2 + x_6 b_2 + a_1 a_2 + a_1 b_2 + a_2 b_2 - a_1 - a_2 - b_2 - x_6 + 1 \\
&+ x_7 a_1 + x_7 b_1 + x_7 b_2 + a_1 b_1 + a_1 b_2 + b_1 b_2 - a_1 - b_1 - b_2 - x_7 + 1 \\
&- a_1 a_2 - a_1 b_1 - a_1 b_2 - a_1 x_1 + a_1
\end{aligned} \tag{15}$$

For all $a_1, a_2, b_1, b_2, \in \{0,1\}$, $a_1 a_2 b_1 b_2$ can be transformed into a combination of linear and quadratic terms by adding new seven qubits, $x_1, x_2, \cdots, x_7$, for each quartic term.

### 2.4. HUBO model with the range dependent Hamiltonian algorithm

Recently, the range dependent Hamiltonian algorithm was proposed[15]. This algorithm divides the domain into subregions that can be represented by the desired number of qubits. Applying this algorithm, $p$ and $q$ can be expressed as follows:

$$p \approx \sum_{l=0}^{n-1} 2^l q_l + S_i \text{ and } q \approx \sum_{l=0}^{n-1} 2^l q_{n+l} + S_j \tag{16}$$

where $S_k = k 2^n$ and $k$ is an integer.

To derive a HUBO model, we insert Eq. (16) into Eq. (6). This yields the summation terms of the first term in Eq. (6), as indicated below:

$$p^2 q^2 - S_i^2 S_j^2 = \left( \sum_{l=0}^{n-1} 2^l q_l + S_i \right)^2 \left( \sum_{l=0}^{n-1} 2^l q_{n+l} + S_j \right)^2 - S_i^2 S_j^2 \tag{17}$$

$$= \left( \sum_{l=0}^{n-1} (2^{2l} + 2^{l+1} S_i) q_l + \sum_{l_1 < l_2} 2^{l_1 + l_2 + 1} q_{l_1} q_{l_2} + S_i^2 \right) \left( \sum_{l=0}^{n-1} (2^{2l} + 2^{l+1} S_j) q_{n+l} \right. \tag{18}$$

$$\left. + \sum_{l_1 < l_2} 2^{l_1 + l_2 + 1} q_{n+l_1} q_{n+l_2} + S_j^2 \right) - S_i^2 S_j^2$$

$$= \sum_{l=0}^{n-1} \left( (2^{2l} + 2^{l+1} S_i) S_j^2 q_l + (2^{2l} + 2^{l+1} S_j) S_i^2 q_{n+l} \right) + \sum_{l=0}^{n-1} \left( 2^{l_1 + l_2 + 1} S_j^2 q_{l_1} q_{l_2} + 2^{l_1 + l_2 + 1} S_i^2 q_{n+l_1} q_{n+l_2} \right) \tag{19}$$

$$+ \sum_{l_1=0}^{n-1} \sum_{l_2=0}^{n-1} \left( 2^{2(l_1 + l_2)} + 2^{l_1 + 2l_2 + 1} S_i + 2^{2l_1 + l_2 + 1} S_j + 2^{l_1 + l_2 + 2} S_i S_j \right) q_{l_1} q_{n+l_2}$$

$$+ \sum_{l_1=0}^{n-1} \sum_{l_2 < l_3} \left( 2^{l_2 + l_3 + 1} (2^{2l_1} + 2^{l_1+1} S_j) \right) q_{l_2} q_{l_3} q_{n+l_1} + \sum_{l_1=0}^{n-1} \sum_{l_2 < l_3} \left( 2^{l_2 + l_3 + 1} (2^{2l_1} + 2^{l_1+1} S_i) \right) q_{l_1} q_{n+l_2} q_{n+l_3}$$

$$+ \sum_{l_1 < l_2} \sum_{l_3 < l_4} 2^{l_1 + l_2 + l_3 + l_4 + 2} q_{l_1} q_{l_2} q_{n+l_3} q_{n+l_4}$$

In Eq. (8), since $(q_l)^2 = q_l$, equation (9) can be obtained. The second term of equation (6) is calculated as follows:

$$-2N(pq - S_i S_j) = -2N\left(\sum_{l=0}^{n-1} 2^l q_l + S_i\right)\left(\sum_{l=0}^{n-1} 2^l q_{n+l} + S_j\right) + 2NS_i S_j \tag{20}$$

$$= -\sum_{l=0}^{n-1} 2^{l+1} N(S_j q_l + S_i q_{n+l}) - \sum_{l_1=0}^{n-1}\sum_{l_2=0}^{n-1} 2^{l_1+l_2+1} N q_{l_1} q_{n+l_2} \tag{21}$$

The HUBO model consists of the sum of Eqs. (19) and (21). And the global minimum energy we need to get is $-N^2 - S_i^2 S_j^2 + 2NS_i S_j$.

### 2.5. HUBO model with the qubit decomposition algorithm

Jun proposed the qubit decomposition algorithm [16]. This algorithm is a method of finding Eq. (7) by continuously using $p$ and $q$ for a specific number of qubits. The advantage of this algorithm is that it can find a given equation with one qubit. We show here how $p$ and $q$ each find a solution with one qubit. This algorithm first finds the minimum energy from the highest order qubit for an exponent of 2. The next step is to find the minimum energy at each step by lowering the order one by one. The larger the energy difference for each step of a given problem, the easier it is to find a solution.

We will explain this algorithm with an example. To factorize 15, let $(p, q)$ be $(p, q) \approx (q_1 + 2q_2 + 4q_3, q_4 + 2q_5 + 4q_6)$. First, we represent to be $(P_1, Q_1) \approx (4q_3, 4q_6)$ with two qubits to find a solution. We can obtain $(P_1, Q_1)$ as $(4,4)$ using the HUBO model. In the next step, we can set $(P_2, Q_2)$ to $(2q_2^+ - 2q_2^-, 2q_4^+ - 2q_4^-)$. We can obtain $(P_2, Q_2)$ as $(0,0)$ using the new HUBO model. Finally, we can set $(P_3, Q_3)$ to $(q_1^+ - q_1^-, q_3^+ - q_3^-)$. Also, we can obtain $(P_3, Q_3)$ as $(-1,1)$ using the new HUBO model. The solution $(p, q)$ is $(P_1 + P_2 + P_3, Q_1 + Q_2 + Q_3)$.

### 3. Experiments

We use a solver, dimod.ExactPolySolver().sample_hubo(), in the D-Wave Ocean SDK to test the HUBO model for the linear systems. This solver represents the energy for every possible number of cases each qubit can have. We tested it on 100 different numbers belonging to the first thousand prime numbers. In all cases, we were able to find pairs of $pq$ and $qp$ using the new HUBO model. The largest number we've tested is 102,454,763. The solver factored the number into two prime numbers, 10,111 and 10,133. The pseudo-code used in the test is shown in Algorithm 1.

---

**Algorithm 1** The HUBO model for prime factorization

---

Determine the number of qubits to be used for each prime number: NQ

Represent $p$ and $q$ as the sum of qubits:

$$(p, q) \approx \left(\sum_{l=0}^{NQ-1} 2^l q_l, \sum_{l=0}^{NQ-1} 2^l q_{NQ+l}\right)$$

Global minimum energy for solution: $-N^2$

▷ Quadratic Terms in Eqs. (10) and (12)

**for** $l_2 = 0: NQ - 1$ **do**

    **for** $l_1 = 0: NQ - 1$ **do**

        $Q(l_1, NQ + l_2) \leftarrow 2^{2(l_1+l_2)} - 2^{l_1+l_2+1} N$

▷ Cubic Terms in Eq.(10)

**for** $l_3 = 0: NQ - 1$ **do**

    **for** $l_1 = 0: NQ - 2$ **do**
        **for** $l_2 = l_1 + 1: NQ - 1$ **do**

$$Q(l_1, l_2, NQ + l_3) \leftarrow 2^{l_1+l_2+2l_3+1}$$

$$Q(l_3, NQ + l_1, NQ + l_2) \leftarrow 2^{l_1+l_2+2l_3+1}$$

▷ Quartic Terms in Eq. (10)

**for** $l_3 = 0: NQ - 2$ **do**

    **for** $l_4 = l_1 + 1: NQ - 1$ **do**
        **for** $l_1 = 0: NQ - 2$ **do**

            **for** $l_2 = l_3 + 1: NQ - 1$ **do**
                $Q(l_1, l_2, NQ + l_3, NQ + l_4) \leftarrow 2^{l_1+l_2+l_3+l_4+2}$

---

        To calculate the first quadratic terms, we multiply $2^{l_1}$ and $2^{l_2}$ and then square them for $n^2$ loops. The total number of flops is $2(n + 1)^2$. In a similar way, the number of $3(n + 1)^2$ flops can be obtained for the second quadratic terms. The total number of flops for quadratic terms is $5(n + 1)^2$. We can consider the change in terms of the first quadratic terms. $2^{2(l_1+l_2)}$ varies from $4^0$ to $4^{2n-2}$. We can represent all of these terms as a sum of $2n + 1$. Rather than directly calculating each coefficient, if we find the amount of change and put it into the coefficient appropriately, we can reduce the amount of calculation by about the square root. In a similar way, calculating the number of flops as a change amount for the cubic term and the quartic term requires $3n + 1$ and $4n + 1$, respectively. We can reduce the amount of calculation once more here. The coefficients of the first quadratic terms and cubic terms are included in the coefficients of quartic terms. Therefore, the total optimized number of flops for the HUBO model is $(4n + 3) + (2n + 3)$.

        Both cubic and quartic terms of this HUBO model have positive coefficients. Therefore, the QUBO model for the factorization can be formulated using Eqs. (13) and (15). The pseudo-code is shown in Algorithm 2. In Algorithm 2, the cubic terms are divided into linear terms, quadratic terms, and constant terms by Eq. (13). The constant term generated here is not included in the QUBO model and appears in the reduced form of the global minimum energy.

---

**Algorithm 2** The QUBO model for prime factorization

Determine the number of qubits to be used for each prime number: NQ

Represent $p$ and $q$ as the sum of qubits:

$$(p, q) \approx \left( \sum_{l=0}^{NQ-1} 2^l q_l, \sum_{l=0}^{NQ-1} 2^l q_{NQ+l} \right)$$

Initial global minimum energy for solution: $GME = -N^2$

▷ Quadratic Terms in Eqs. (10) and (12)

**for** $l_2 = 0: NQ - 1$ **do**

    **for** $l_1 = 0: NQ - 1$ **do**

        $Q(l_1, NQ + l_2) \leftarrow 2^{2(l_1+l_2)} - 2^{l_1+l_2+1}N$

▷ Cubic Terms in Eq.(10)

np = $2NQ - 1$

**for** $l_3 = 0: NQ - 1$ **do**

    **for** $l_1 = 0: NQ - 2$ **do**

        **for** $l_2 = l_1 + 1: NQ - 1$ **do**

            np ← np+1

            Reduction algorithm: Two cubic terms to the combination of linear and quadratic terms

            $Q(l_1, l_1) \leftarrow -2^{l_1+l_2+2l_3+1}$

            $Q(l_2, l_2) \leftarrow -2^{l_1+l_2+2l_3+1}$

            $Q(l_3, l_3) \leftarrow -2^{l_1+l_2+2l_3+1}$

            $Q(l_1 + NQ, l_1 + NQ) \leftarrow -2^{l_1+l_2+2l_3+1}$

            $Q(l_2 + NQ, l_2 + NQ) \leftarrow -2^{l_1+l_2+2l_3+1}$

            $Q(l_3 + NQ, l_3 + NQ) \leftarrow -2^{l_1+l_2+2l_3+1}$

            $Q(l_{np}, l_{np}) \leftarrow -2^{l_1+l_2+2l_3+1}$

            $Q(l_{np} + 1, l_{np} + 1) \leftarrow -2^{l_1+l_2+2l_3+1}$

            $Q(l_1, l_2) \leftarrow 2^{l_1+l_2+2l_3+1}$

            $Q(l_1, l_3 + NQ) \leftarrow 2^{l_1+l_2+2l_3+1}$

            $Q(l_2, l_3 + NQ) \leftarrow 2^{l_1+l_2+2l_3+1}$

            $Q(l_3, l_1 + NQ) \leftarrow 2^{l_1+l_2+2l_3+1}$

            $Q(l_3, l_2 + NQ) \leftarrow 2^{l_1+l_2+2l_3+1}$

            $Q(l_1 + NQ, l_2 + NQ) \leftarrow 2^{l_1+l_2+2l_3+1}$

            $Q(l_1, l_{np}) \leftarrow 2^{l_1+l_2+2l_3+1}$

            $Q(l_2, l_{np}) \leftarrow 2^{l_1+l_2+2l_3+1}$

            $Q(l_3, l_{np} + 1) \leftarrow 2^{l_1+l_2+2l_3+1}$

            $Q(l_3, l_{np} + 1) \leftarrow 2^{l_1+l_2+2l_3+1}$

            $Q(l_1 + NQ, l_{np} + 1) \leftarrow 2^{l_1+l_2+2l_3+1}$

            $Q(l_2 + NQ, l_{np} + 1) \leftarrow 2^{l_1+l_2+2l_3+1}$

            $GME = GME - 2^{l_1+l_2+2l_3+2}$

np ← np−5

▷ Quartic Terms in Eq. (10)

**for** $l_3 = 0: NQ - 2$ **do**

```
for l_4 = l_1 + 1: NQ − 1  do
    for l_1 = 0: NQ − 2 do
        for l_2 = l_3 + 1: NQ − 1   do
            np ← np+7
            do Reduction algorithm two steps
            GME = GME − 3 ∗ 2^{l_1+l_2+l_3+l_4+3}
```

---

Two cubic terms with positive coefficients can be formulated into 10 linear and quadratic terms using Eq. (13). In the process of converting the two cubic terms into the form of the QUBO model, two new qubits are added. Eq. (15) is used to formulate each quartic term in the form of a QUBO model. First, Eq. (13) is applied for three qubits. Then, the remaining qubit is multiplied by the newly created terms. Here, Eq. (15) is formulated by applying Eq. (13) again to the newly created cubic terms. By formulating cubic and quartic terms into linear and quadratic terms, the QUBO model for prime factorization can be obtained.

We prime factorize the natural number N by two prime numbers, p and q, to test the QUBO model for prime factorization. Figure 1 is the result obtained using a qpu solver for the QUBO model. In the process of calculating the HUBO model as the QUBO model, different energies appear for the same solution because new logical qubits are used. The energy we can see as a solution is the global minimum energy of -2756.

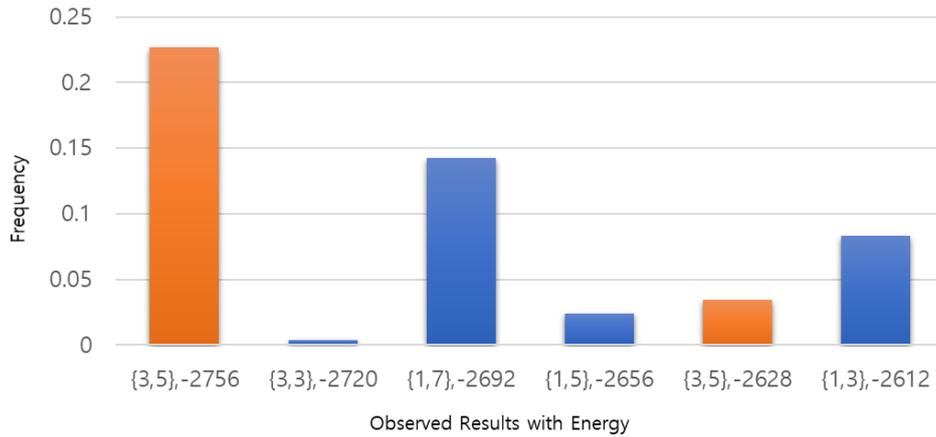

Figure 1. Experimental result of the QUBO model on D-Wave machine. The x-axis represents solution and energy, and the y-axis represents frequency. The solution was obtained with a qpu solver using 15 logical qubits.

Finally, we applied the range dependent algorithm to the HUBO model to factorize 1,000,070,001,221. The pseudo-code used here can be seen in Algorithm 3. The HUBO model to which this algorithm is applied shows the global minimum energy $-4,900,170,941,490,841$ only in the section where the solution exists. To solver the HUBO model with range dependent algorithm, we used 12 logical qubits, and each $S_i$ and $S_j$ is 1,000,000.

---

**Algorithm 3** The HUBO model with range dependent Hamiltonian algorithm for prime factorization
---

Determine the number of qubits to be used for each prime number: NQ

Represent $p$ and $q$ as the sum of qubits:

$$(p, q) \approx \left( \sum_{l=0}^{NQ-1} 2^l q_l + S_i, \sum_{l=0}^{NQ-1} 2^l q_{NQ+l} + S_j \right)$$

Global minimum energy for solution: $-N^2 - S_i^2 S_j^2 + 2NS_i S_j$

**for** each $S_i$ and $S_j$ **do**

    **for** $l = 0: NQ - 1$ **do**

        $Q(l, l) \leftarrow (2^{2l} + 2^{l+1} S_i) S_j^2 - 2NS_j 2^l$

        $Q(NQ + l, NQ + l) \leftarrow (2^{2l} + 2^{l+1} S_j) S_i^2 - 2NS_i 2^l$

    **for** $l_1 = 0: NQ - 2$ **do**

        **for** $l_2 = l_1 + 1: NQ - 1$ **do**

            $Q(l_1, l_2) \leftarrow 2^{l_1+l_2+1} S_j^2$

            $Q(NQ + l_1, NQ + l_2) \leftarrow 2^{l_1+l_2+1} S_i^2$

    **for** $l_1 = 0: NQ - 1$ **do**

        **for** $l_2 = 0: NQ - 1$ **do**

            $Q(l_1, NQ + l_2) \leftarrow 2^{2(l_1+l_2)} + 2^{l_1+2l_2+1} S_i + 2^{2l_1+l_2+1} S_j + 2^{l_1+l_2+2} S_i S_j - N2^{i+j+1}$

    **for** $l_3 = 0: NQ - 1$ **do**

        **for** $l_1 = 0: NQ - 2$ **do**

            **for** $l_2 = l_1 + 1: NQ - 1$ **do**

                $Q(l_2, l_3, NQ + l_1) \leftarrow 2^{l_2+l_3+1}(2^{2l_1} + 2^{l_1+1} S_j)$

                $Q(l_1, NQ + l_2, NQ + l_3) \leftarrow 2^{l_2+l_3+1}(2^{2l_1} + 2^{l_1+1} S_i)$

    **for** $l_3 = 0: NQ - 2$ **do**

        **for** $l_4 = l_1 + 1: NQ - 1$ **do**

            **for** $l_1 = 0: NQ - 2$ **do**

                **for** $l_2 = l_3 + 1: NQ - 1$ **do**

                      $Q(l_1, l_2, NQ + l_3, NQ + l_4) \leftarrow 2^{l_1+l_2+l_3+l_4+2}$

---

## 4. Discussion

    The new algorithm for prime factorization has cubic and quartic terms, so it is suitable for the HUBO model. In this paper, we factored two prime numbers for 102,454,763, but larger numbers are possible. The reason we used 26 logical qubits in the prime factorization problem is to prevent the capacity from becoming too large when receiving results from the D-Wave system. It is possible to convert the HUBO model to the QUBO model using Algorithm 2, but this requires a large number of additional qubits. We tested an example of $N = 15$ for the QUBO model. We used $(p, q) = (1 + 2q_1 + 4q_2, 1 + 2q_3 + 4q_4)$ to solve this problem. Among the additional 11 logical qubits used, 4 qubits were used when converting cubic terms into the form of the QUBO model. The remaining 7 qubits were used in the process of converting the quartic term into the QUBO model. In Fig. 1, we found that different energies appear for the same solution $(p, q)$. This is because additional qubits were used in the process of calculating the HUBO model as the QUBO model. It doesn't matter because we can find the solution when the global minimum energy appears. Seven additional logical qubits are required to transform each quartic term into the terms of a QUBO model. If each $p$ and $q$ uses $n$ logical qubits, our QUBO model requires $\frac{7n^2(n-1)^2}{4}$ additional qubits for quartic terms. Therefore, our QUBO model in the current quantum annealer system is not suitable for prime factorization of large natural numbers. If $q_i q_j q_l q_m$ can be implemented in a quantum annealer system as the concept of a delta measure $\delta_{ijlm}$, where $\delta_{ijlm}$ is 1 when $i = j = l = m = 1$ and $\delta_{ijlm}$ is zero for the other case, it is expected that there will be tremendous progress in the development of the QUBO

model.

We applied the range dependent Hamiltonian algorithm to the HUBO model. This algorithm divides the domain into several smaller intervals. The advantage of this algorithm is that it can find a solution if it can be expressed in qubits only for small intervals of change. In addition, since it can be calculated in each independent subrange, like the domain division of parallel computing, the more the hardware develops, the better it is applied. We apply this algorithm to factorize 1,000,070,001,221 into 1,000,033 and 1,000,037. We used 12 logical qubits when factoring this value, and each $S_i$ and $S_j$ used 1,000,000. In general, $S_i$ and $S_j$ can be expressed as $k2^n$, but we used $10^n$ to make calculations easier. The number 1 trillion is close to the largest number that can be factored through the D-Wave system. When calculating a number greater than 5% of the number we calculated, an error message was sent because the coefficient of each term of the HUBO model was not allocated normally in the D-Wave system. We used the Decimal function to solve this problem, but it was not properly allocated in the D-Wave system. As another approach to the large number $N$, we calculated the double type for each coefficient, but could not find the global minimum energy in the D-Wave system due to floating-point error mitigation. If the number of qubits in the hybrid solver increases and the coefficients can be accurately expressed in large numbers, we think that the prime factorization problem will be solved.

**Declaration of interests**

The authors have no competing interests which may have influenced the work shown in this manuscript.

**Acknowledgement**

This research was supported by the quantum computing technology development program of the National Research Foundation of Korea (NRF) funded by the Korean government (Ministry of Science and ICT (MSIT)) (No. 2020M3H3A111036513 and 2019R1A6A1A11051177). The authors would like to thank Dr. Lee for summarizing the existing prime factorization problem.

Different Columns." *Frontiers in Physics* (2022): 518.

13. D-Wave Systems, Inc. D-Wave Ocean Software Documentation, [Online]. Available: https://docs.ocean.dwavesys.com
14. O'Malley, Daniel, and Velimir V. Vesselinov. "Toq. jl: A high-level programming language for d-wave machines based on julia." *2016 IEEE High Performance Extreme Computing Conference (HPEC)*. IEEE, 2016.
15. Hyunju Lee, and Kyungtaek Jun. "Range dependent Hamiltonian Algorithm for numerical QUBO formulation*." arXiv preprint* arXiv:2202.07692 (2022).
16. Kyungtaek Jun, Qubits decomposition for highly precision calculation, electronic resource: python source code: https://github.com/ktfriends/Numerical_Quantum_Computing/tree/main/HP (2022)